\documentclass[11pt]{article}
\usepackage{paithan}
\usepackage{setspace}

\title{A PSPACE-complete Graph Nim}
\author{Kyle G. Burke \\ Colby College \\ paithanq@gmail.com \and Olivia C. George\\ \phantom{oliviacgeorge@yahoo.com} \\ \phantom{monkey}}

\begin{document}
 
 \maketitle
 
 \begin{abstract}
  
  We build off the game \ruleSet{NimG} \cite{StockmanREU:2004} to create a version named \ruleSet{Neighboring Nim}.  By reducing from \ruleSet{Geography}, we show that this game is \complexC{PSPACE}-hard.  The games created by the reduction share strong similarities with \ruleSet{Undirected (Vertex) Geography} and regular \ruleSet{Nim}, though these are both solvable in polynomial-time.  This application of graphs can be used as a form of game sum with any rulesets, not only \ruleSet{Nim}.
  
  \end{abstract}
  
  \section{Background}
  
  
  
  \subsection{Algorithmic Combinatorial Game Theory}
  
  Most of the results here revolve around the computational complexity of determining which player has a winning strategy from a given game position.  There exist faster algorithms to solve this problem for some rulesets than for others.  For each ruleset, we consider the computational problem that could be solved by such an algorithm.  We will refer to both the ruleset and problem by the same name.
  
  We strongly encourage readers unfamiliar with these topics to refer to \cite{AlgGameTheory_GONC3}.

  \subsection{Terminology}
  
%
%
%

There is a small amount of non-standard terminology used in this paper.
  
\begin{itemize}

  \item We use the word \emph{sticks} to refer to the objects in nim heaps.  Thus, a nim heap of size six contains six sticks.

  \item An \emph{optimal sequence set} is a set of sequences of plays for both players such that any move deviating from one of the sequences results in an \fancy{N}-position.  No move in that sequence should be non-optimal for either player.  Thus, if a player does not know whether they have a winning strategy, adhering to an optimal sequence is at least as good as any other move.
\end{itemize}
  
  \subsection{\ruleSet{Nim}}
  
  \ruleSet{Nim} is an impartial game played on a collection of heaps, each with a non-negative number of sticks.  On a player's turn, they choose a non-empty pile and remove as many sticks as desired (at least one) from that pile.  A player loses when they cannot remove sticks (all piles are empty).
  
  \ruleSet{Nim} is a classic impartial game, being the basis of Nimbers and Sprague-Grundy theory\cite{Sprague:1936}\cite{Grundy:1939}.  \ruleSet{Nim} has lots of nice properties, from easy evaluation of games to obvious composition of two \ruleSet{Nim} games (the sum is just a new \ruleSet{Nim} game).
  
  \subsection{\ruleSet{NimG}}
  
  \ruleSet{Nim} has been extended to incorporate graphs so that nim heaps are assigned to either edges or vertices.  There are three different versions of the game named \ruleSet{NimG}.  In all three versions, a turn consists of both traversing an edge of the graph and removing sticks from a visited element.
  
  \subsubsection{Edge-heap NimG}
  
  Fukuyama describes \ruleSet{NimG} where nim heaps are embedded into the edges of the graph\cite{DBLP:journals/tcs/Fukuyama03}.  On each turn, the current player chooses an edge to traverse (which has at least 1 stick on it) and removes any number of sticks from that edge.  The next player then starts on the vertex on the other end of that edge and must choose an adjacent edge for their move.  When there are no more edges with sticks adjacent to the current vertex, the current player loses.  Many results for this game are known on complete graphs\cite{Erickson:nimOnCompleteGraph}.
  
  \subsubsection{Vertex-heap NimG}
  
  In \ruleSet{Vertex NimG}, players similarly move from one vertex to another, but heaps are connected to the vertices instead of edges\cite{StockmanREU:2004}.  The two variants can be easily described here as: ``remove sticks, then move'' and ``move, then remove sticks''.  In both cases, a player loses if they cannot complete their turn.  The main topic of this paper is a variant of ``move, then remove''.
  
  \subsection{Geography}
  
  We will use \ruleSet{Geography} to show the \complexC{PSPACE}-hardness of \ruleSet{Neighboring Nim}.  There are many flavors of \ruleSet{Geography}; we use the term to refer to \ruleSet{Directed Vertex Geography}.  This impartial game is played on a directed graph; each turn begins with a vertex chosen.  The current player's turn consists of choosing an outgoing arc that leads to a vertex that hasn't been visited yet.  The next player then starts their turn with the resulting vertex selected.  We formally describe the ruleset as follows:
  
  \begin{Definition}[Geography]
    (Directed Vertex) Geography positions are described by: $G = (V, E)$ and $v \in V$.  Move options for $(G, v)$ are all $(G', v')$ where:
    \begin{itemize}
      \item $(v, v') \in E$,
      \item $V' = V \setminus \{v\}$,
      \item $E'$ is the subset of $E$ induced by $V'$, and
      \item $G' = (V', E')$.
    \end{itemize}

  \end{Definition}

  \ruleSet{Geography} is known to be \complexC{PSPACE}-complete\cite{PapadimitriouBook:1994}.
  
  \section{\ruleSet{Neighboring Nim}}
  
  We define the ruleset \ruleSet{Neighboring Nim} to be similar to the ``move then remove'' version of \ruleSet{NimG}, but also allow players to choose to play on the same vertex as the last move as though each vertex has a self-loop.  Note that standard \ruleSet{Nim} is equivalent to a game of \ruleSet{Neighboring Nim} on a complete graph with each heap on a separate vertex.  A more formal definition follows.
  
\begin{Definition}[\ruleSet{Neighboring Nim}]
  \ruleSet{Neighboring Nim} positions are described by $G = (V, E)$, $w: V \rightarrow \mathbb{N}$, and $x \in V$.  The options for $(G, w, x)$ are all $(G, w', x')$ where $w': V \rightarrow \mathbb{N}$ and
  \begin{itemize}
    \item $x' = x$ or $\{x, x'\} \in E$,
    \item $w'(x') < w(x')$, and
    \item $\forall v \in V \setminus \{x'\}: w'(v) = w(v)$.
  \end{itemize}
  
 \end{Definition}
 
 Our main result for this paper is that \ruleSet{Neighboring Nim} is \complexC{PSPACE}-hard.  Since our analysis uses graphs with a small number of sticks on each vertex, we define a version of the game with a bounded number of sticks per vertex.
 
\begin{Definition}[\ruleSet{$k$-Neighboring Nim}]
$\phantom{y}\\$
\emph{\ruleSet{$k$-Neighboring Nim}} is the same ruleset as \ruleSet{Neighboring Nim}, except that the weight function, $w$, has bounded range: $[0, k]$.
\end{Definition}

We are able to show that \ruleSet{2-Neighboring Nim} is \complexC{PSPACE}-complete, and thus \ruleSet{$c$-Neighboring Nim} is also \complexC{PSPACE}-complete for any constant $c \geq 2$.  The case for \ruleSet{1-Neighboring Nim} is solvable in polynomial time, since this game is equivalent to \ruleSet{Undirected (Vertex) Geography} \cite{DBLP:journals/tcs/FraenkelSU93}.  Thus, if $\complexC{P} \neq \complexC{PSPACE}$, allowing a second stick on some vertex-heaps can greatly increase the computational hardness of determining the winning player!

\section{Computational Complexity of \ruleSet{Neighboring Nim}}
\label{section:hardness}

\subsection{\complexC{PSPACE}-hardness}
  
The following is the main result of this paper.
  
\begin{theorem}[Hardness]
  \label{theorem:hardness}
 \ruleSet{Neighboring Nim} is \complexC{PSPACE}-hard.
\end{theorem}
 
 We will show the hardness of this problem by reducing from the game \ruleSet{Geography}, which is \complexC{PSPACE}-hard\cite{LichtensteinSipser:1980}.
 
\begin{Proof}
 Given any \ruleSet{Geography} position, we will give an algorithm to construct an equivalent \ruleSet{Neighboring Nim} state, meaning that there is a win in the \ruleSet{Geography} position exactly when there is a win in corresponding \ruleSet{Neighboring Nim} position.  First we will describe the method for generating these positions, then prove their equivalence.

 Let $GG$ be a \ruleSet{Geography} position on the directed and unweighted graph $G = (V,E)$.  We define a new undirected graph, $G' = (V', E')$ with weights on the vertices $w: V' \rightarrow \mathbb{N}$ in the following way.  $\forall v \in V:$ let $X_{v} \in V'$ and set $w(X) = 1$.  Also, $\forall (y, z) \in E:$ (edge directed from $y$ to $z$) let $a_{y,z}, b_{y,z}, c_{y,z}, d_{y,z}, e_{y,z}, f_{y,z}, g_{y,z} \in V'$ where, ignoring the $(y,z)$-subscripts, $w(a) = w(b) = w(c) = w(e) = w(f) = w(g) = 1$, $w(d) = 2$ and, 

$(X_{y},a), (a,b), (b,c), (c,d), (b,e), (e,f), (d,f), (d,g), (f,g), (g,X_{z}) \in E'$.

See Figure \ref{fig:edgeReduction} for a visual description.
 
\begin{figure}[h]
 \centering
 \includegraphics[scale = .6]{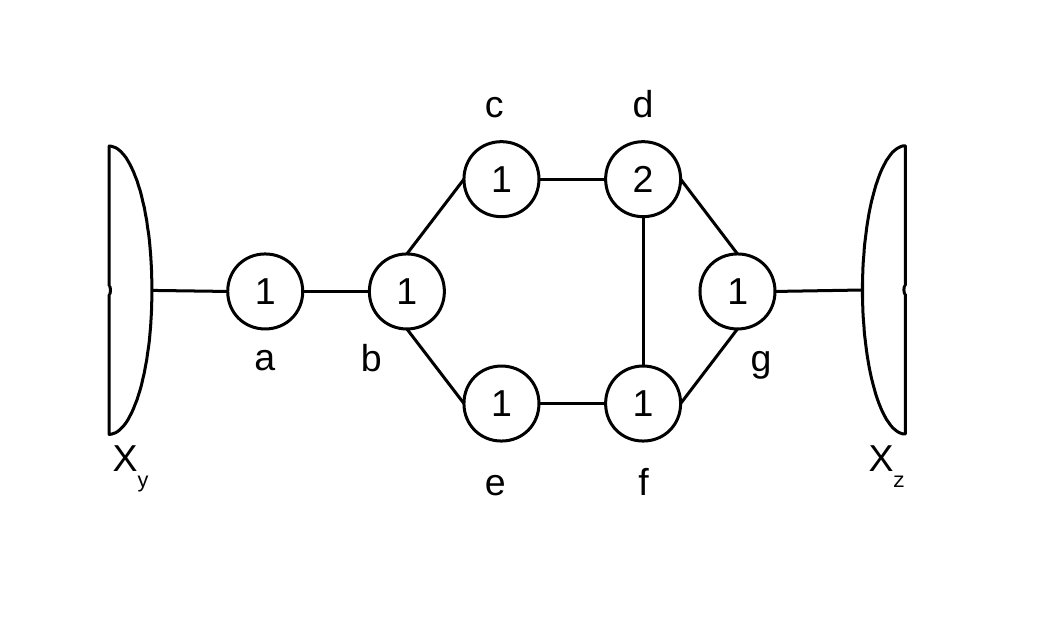}
 \caption{Our main gadget: reduce each directed edge from $y$ to $z$ to the undirected weighted graph shown here.}
 \label{fig:edgeReduction} 
\end{figure}
  
 The resulting $G'$ is the graph for our \ruleSet{Neighboring Nim} position equivalent to $GG$.  The only final step is to declare that if $GG$ has a starting vertex, $v$, then $X_{v} \in V'$ is the starting vertex (where the previous play had been made) in our game and $w(X_{v})$ is set to 0 instead of 1.  
 
 To complete the reduction, we must show that the structure in Figure \ref{fig:edgeReduction} ``acts'' like a directed edge in \ruleSet{Geography}.  Thus, we must prove:

  \begin{itemize}
 \item  Moving ``backwards'' is a losing play.  If the previous play was at $X_{z}$, then a backwards play would be to remove the only stick at $g_{(y,z)}$.  A backwards play results in an $\fancy{N}$-position.

 \item  The same player moving into the gadget should also move out.  If a player moves from $X_y$ to $a_{(y, z)}$, then in an optimal sequence of plays, the same player will move from $g_{(y,z)}$ to $X_z$.
\end{itemize}

 We prove the former in Lemma \ref{lemma:dontGoBackwards} and the latter in Lemma \ref{lemma:stickToTheScript}.  The result is that each of these gadgets (as in Figure \ref{fig:edgeReduction}) in the \ruleSet{Neighboring Nim} position works just like a (directed) edge in \ruleSet{Geography}.  Trying to go backwards will result in losing and, if players play optimally, they both might as well continue through each gadget.  

 
\begin{Lemma}[Don't Go Backwards]
 \label{lemma:dontGoBackwards}
 Any play from $(X_{z})$ to $g_{(y,z)}$ (for all $y$) results in an $\fancy{N}$-position.
\end{Lemma}

(See Appendix \ref{appendix:dontGoBackwards} for a proof of this claim.)  This implies that our gadgets are directed: if a player tries to go ``backwards'': from an $X$-vertex to an $i$-vertex, the opponent will have a winning strategy.

To finish showing that our gadget acts like a directed edge, we must prove that ``nothing can go wrong'' during a regular forward traversal of the structure.  To this end, we find two sequences that constitute an optimal sequence set through the gadget, thus showing that neither player benefits from deviating from the sequence.  In order to get from one end of the gadget (as in Figure \ref{fig:edgeReduction}) to the other, the following sequence of moves suffices (let Alice and Bob be our two players; we will again ignore subscripts): Alice ``takes'' $a$, Bob takes $b$, Alice takes $e$, Bob takes $f$, Alice decrements $d$ by 1, Bob takes $g$, Alice takes $X_z$.  Note that the same player (in this example, Alice) who chooses to take $a$ also moves to $X_z$.  The other sequence is where Bob takes $c$ instead of $g$---here Alice will take the remaining object at $d$ and Bob will be forced to take $g$, rejoining with the first sequence.  See Figure \ref{fig:safeSequence} for a visual description of the safe sequences.  We must prove that neither player benefits from deviating from these sequences.  To do this, we show that any deviation is a losing move.

 \begin{figure}[h]
\centering
 \includegraphics[scale = .6]{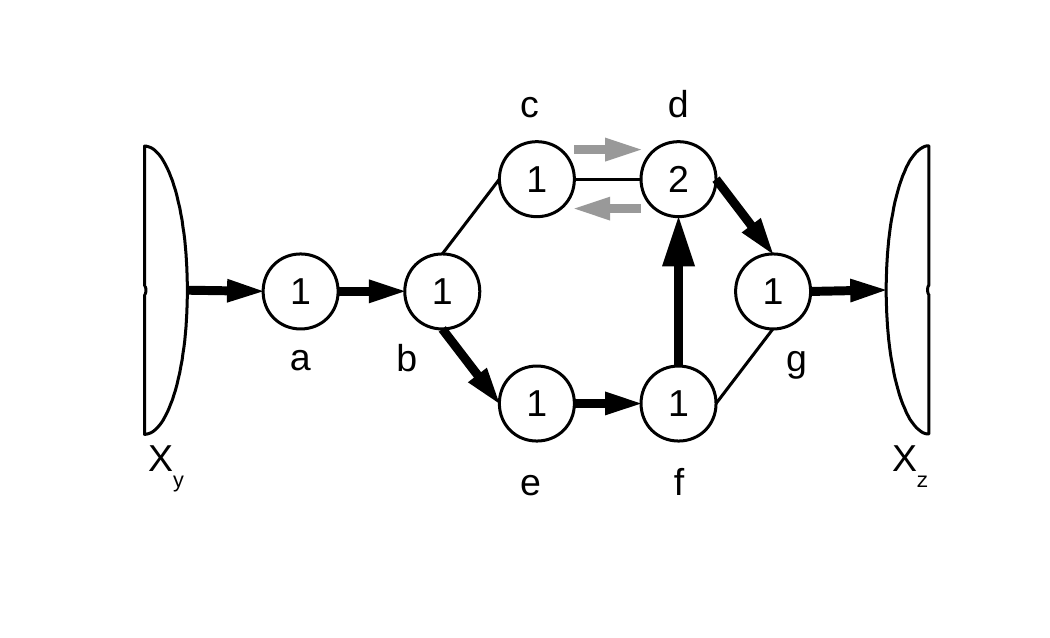}
 \caption{This sequence of moves is ``safe'' for both players to traverse the gadget.  The two gray arrows indicate the potential additional moves in the second sequence.  Each move assumes that exactly one object is taken from a vertex.}
 \label{fig:safeSequence} 
\end{figure}
  
\begin{Lemma}[Stick to the Script]
 \label{lemma:stickToTheScript}
 Let the notation $k(p)$ denote taking $p$ objects from vertex $k$ in a turn.  Then, after the plays $(\ldots, X_y(1), a(1))$ any play deviating from the following sequences is a losing move: \\
$(b(1), e(1), f(1), d(1), g(1), X_z(1))$\\
 $(b(1), e(1), f(1), d(1), c(1), d(1), g(1), X_z(1))$.
\end{Lemma}

(See Appendix \ref{appendix:stickToTheScript} for the proof of this claim.)  This implies that once a player makes an appropraite move onto the gadget (playing on an $a$-node) any ``safe'' sequence of moves in the gadget results in that same player making the play at the opposite $X$ node.  The two above claims combined show that our gadget correctly models a directed edge in a graph just between the $X$ nodes.

Thus, for any edge $(y,z)$ in our \ruleSet{Geography} position, $GG$, the move to $a_{(y,z)}$ will result in the same player moving to $X_z$ as desired.  Also, since we proved players shouldn't go backwards, this game is equivalent to $GG$; the first player has a winning strategy in $GG$ exactly when the first player has a winning strategy in this \ruleSet{Neighboring Nim} position.
 
Thus, \ruleSet{Neighboring Nim} is \complexC{PSPACE}-hard.
\end{Proof}

The hardness of \ruleSet{Vertex NimG} follows directly.

\begin{Corollary}[\ruleSet{Vertex NimG} hardness]
  \ruleSet{Vertex NimG} is \complexC{PSPACE}-hard.
\end{Corollary}

\begin{Proof}
  Neighboring Nim is a special case of \ruleSet{Vertex NimG} where all vertices have self-loops.  Thus, \ruleSet{Vertex NimG} is also \complexC{PSPACE}-hard.
\end{Proof}

\subsection{Speculation on Completeness}
  
Unfortunately, \ruleSet{Neighboring Nim} is not automatically \complexC{PSPACE}-complete as games could take a number of moves exponential in the size of the description of the game.  For example, a vertex can have a number of sticks exponential in the amount of bits needed to express that number and the rest of the graph.  We leave this unsolved as Open Problem \ref{open:inPSPACE}.  There are good arguments to conjecture either way.  

On one hand, it seems to not be inside \complexC{PSPACE}.  Games can last an exponential number of turns, so the game trees are extremely tall.  A straight-forward brute-force traversal can't be performed in polynomial space.  

On the other hand, it might be inside \complexC{PSPACE}.  Although there are many \complexC{EXPTIME}-hard rulesets, the authors know only of loopy examples.  This means they can have positions that repeat during the course of a game, which cannot occur in \ruleSet{Neighboring Nim}.  Additionally, Nim heaps are well-understood, perhaps increasing the size of the heaps doesn't greatly increase the difficulty of finding strategies.

\subsection{\complexC{PSPACE}-complete versions}

We can sidestep this problem a bit by using our bounded-heap-size version of the game. 
 
\begin{Corollary}[\ruleSet{2-Neighboring Nim} Completeness] $\phantom{xx}\\$
  \ruleSet{$k$-Neighboring Nim} is \complexC{PSPACE}-complete for any $k \geq 2$.
\end{Corollary}

\begin{Proof}
  The result of the reduction from Theorem \ref{theorem:hardness} is always a \ruleSet{2-Neighboring-Nim} position.  Thus, the \complexC{PSPACE}-hardness holds for this subset of positions as well.  The positions are in \complexC{PSPACE} because $k$ bounds the maximum number of moves per vertex.
\end{Proof}
 
\section{Generalization}
 
This graph-embedding technique works with games other than \ruleSet{Nim}.  Given a graph, assign different game states to the vertices, and use similar rules: players may make one move legal in the game in any vertex neighboring the last play.  We define this formally.

\begin{Definition}[\ruleSet{Neighboring-$R$}]
  Given any ruleset $R$, \ruleSet{Neighboring-$R$} has positions of the form $G = (V, E)$, $w: V \rightarrow positions(R)$, and $x \in V$.  The left options for $(G, w, x)$ are $(G, w', x')$ where $w': V \rightarrow positions(R)$ and:
  \begin{itemize}
    \item $x' = x$ or $\{x, x'\} \in E$,
    \item $w'(x')$ is a left option of $w(x')$, and
    \item $\forall v \in V \setminus \{x'\}: w'(v) = w(v)$.
  \end{itemize}
  The right options are defined analagously: $w'(x')$ must be a right option of $w(x')$.
\end{Definition}

\subsection{Inequivalent Positions}

\begin{figure}[h]
\centering
 \includegraphics[scale = .6]{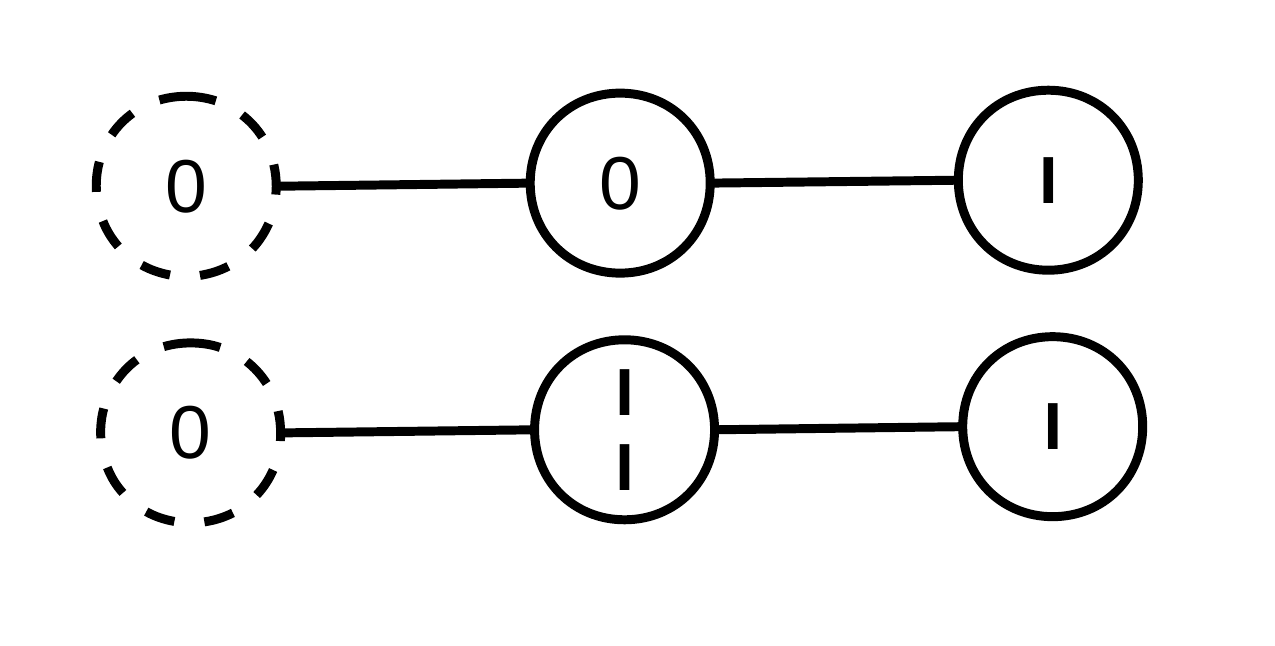} 
 \caption{Two \ruleSet{Neighboring Nim} positions.  In both, the last move was made on the dashed vertex.  The value of the above game is zero, the value of the bottom game is $*$.} 
 \label{fig:inequivalentGraphs} 
\end{figure}

This new definition allows a \ruleSet{Neighboring Nim} vertex to contain multiple heaps instead of only a single heap.  Although each Nim position is equivalent to a single heap, that equivalence doesn't carry over in the Neighboring situation.  Consider the two \ruleSet{Neighboring Nim} games in figure \ref{fig:inequivalentGraphs}.  (The previous move was made on the left-most vertex in both cases.)  The values of the games embedded in the left vertices are both 0, the values on the middle vertices are both 0, and the values of the rightmost games are both $*$.  However, the overall value of the positions are not equivalent.

In the top game, there are no move options, so the value is $0$.  In the bottom position, the next player can move to the middle vertex, even though the value of the Nim game there is also zero.  After that move there are exactly two moves remaining.  Thus, the initial game has exactly three moves remaining and has a value of $*$.

\subsection{Generalized Hardness}

The next result allows us to say something about the hardness of graph-embedded versions of many impartial games.

\begin{theorem}[Neighboring-$R$ Hardness] 
  For any ruleset, $R$, which has positions identical to $0, *$, and $*2$, \ruleSet{Neighboring-$R$} is \complexC{PSPACE}-hard.
\end{theorem}

\begin{Proof}
  Two positions are identical if they have isomorphic game trees.  Replacing the nim heaps of size 1 and 2 with $*$ and $*2$, respectively, in the reduction of theorem \ref{theorem:hardness} doesn't change the winnability of the resulting games.  Thus, the reduction applies to $R$.
\end{Proof}

\section{Conclusions}
 
Building on algorithmic work analyzing different versions on \ruleSet{NimG}, we present \ruleSet{Neighboring Nim}, a new \complexC{PSPACE}-hard game.  
 
An interesting aspect of the hardness of \ruleSet{Neighboring Nim} is the juxtaposition with \ruleSet{Vertex Geography}.  \ruleSet{1-Neighboring Nim} is the same ruleset as \ruleSet{Undirected Vertex Geography}, which is solvable efficiently\cite{DBLP:journals/tcs/FraenkelSU93}.  However, by adding an extra stick to just a few vertices, we can push the game into \complexC{PSPACE}-hardness!

Furthermore, we can replace \ruleSet{Nim} and apply the graph-embedding concept to any other ruleset, $R$, to create \ruleSet{Neighboring-$R$}.  

\section{Future Work}

There are many extensions to the work described here.  The most prominent is certainly the unknown completeness of \ruleSet{Neighboring Nim}.

\begin{OpenProblem}
 \label{open:inPSPACE}
   $\ruleSet{Neighboring Nim} \in \complexC{PSPACE}$?
\end{OpenProblem}

Additionally, there is still work to be done on the computational hardness of other flavors of \ruleSet{NimG}.

\begin{OpenProblem}
  What is the computational complexity of \ruleSet{Edge NimG}?
\end{OpenProblem}

\begin{OpenProblem}
  What is the computational complexity of \ruleSet{Vertex NimG} on graphs without self-loops?
\end{OpenProblem}

Other explorable problems include the hardness of other versions of \ruleSet{Neighboring-$R$}.

\begin{OpenProblem}
  \label{open:equivalentPositionHardness}
  Is \ruleSet{Neighboring-$R$} \complexC{PSPACE}-hard if $R$ includes any positions \emph{equivalent} to $*$ and $*2$?
\end{OpenProblem}

Note that open problem \ref{open:equivalentPositionHardness} is a stronger statement than shown here because equivalent does not necessarily mean identical.

\begin{OpenProblem}
  For which other computationally easy rulesets, $R$, is \ruleSet{Neighboring-$R$} hard? 
\end{OpenProblem}

\begin{OpenProblem}
  Are there strictly partisan positions of a ruleset, $R$, that can be used to show \ruleSet{Neighboring-$R$} is hard?  How small can the game trees be to get a hard game?
\end{OpenProblem}

\section{Acknowledgments}
 
The authors would like to thank those who played \ruleSet{Neighboring Nim} with them, including three excellent Wittenberg University students: Deanna Fink, Dang Mai and Ernie Heyder, as well as Professor Doug Andrews who defeated the first author over and over again.  We would also like to thank Adam Parker for listening to initial versions of the hardness proof and proofreading this paper.

\bibliographystyle{plain}

\appendix

\section{Proof of Lemma \ref{lemma:dontGoBackwards}}
\label{appendix:dontGoBackwards}
 
\begin{Lemma}[Don't Go Backwards]
 Any play from $(X_{z})$ to $g_{(y,z)}$ (for all $y$) is suboptimal.
\end{Lemma}
  
We will refer to the player who moves from $X_z$ to $g$ (we will leave out the subscript for the internal vertices) as the ``foe'' while the other player is the ``hero''.  We will show that the hero has a winning strategy after a backwards move.  We can now look at two cases, each depending on the state of the game outside the gadget.

The first is the case where the move from $a$ to $X_y$ would be a winning play.  In this case, the hero can next move from $g$ to $d$ and take both of the objects there.  The foe has two options, both of which, we show, allow the hero to win.  

\begin{itemize}
\item[1.] \textbf{The foe moves to $c$.} In this case the hero must choose to go to $b$.  The foe can now either choose to move to $a$---in which case the hero will gladly move to $X_y$ and win as we assumed---or to $e$.  Then the hero simply takes the object at $f$ and, as there are no more moves, the hero has won.  

\item[2.] \textbf{The foe moves to $f$.}  The hero must then take $e$ and the foe must take $b$.  The hero can then move to $c$ and win the game.
 
\end{itemize}
 
The second major case assumes that the move from $a$ to $X_y$ is a losing play.  Here, the hero can still move to $d$ (from $g$) but will take only one of the objects.  Now the foe has three options: taking the other object at $d$, moving to $c$ or moving to $f$.  We show all to be losses.
  
 \begin{itemize}
  \item[1.]  \textbf{Foe moves to $c$.}  Now the hero should take the remaining object at $d$.  The following sequence must occur: foe must take $f$, hero at $e$, foe at $b$, hero at $a$, followed by the foe at $X_y$, a losing move by our assumption.
  \item[2.]  \textbf{Foe takes the remaining object at $d$.}  The hero will choose to take $c$, so the foe must take $b$.  The hero can then take $a$, forcing the foe to take $X_y$, a losing move by our assumption.
  \item[3.]  \textbf{Foe takes $f$.}  The hero should then take $e$ so the foe must take $b$.  Again, the hero can take $a$, so the foe must move to $X_y$, a losing move by our assumption.
\end{itemize}
  
 Thus, it is a losing play to move from an $X$-vertex to a $g$-vertex.

\section{Proof of Lemma \ref{lemma:stickToTheScript}}
\label{appendix:stickToTheScript}
  
\begin{Lemma}[Stick to the Script]
  Let the notation $k(p)$ denote taking $p$ objects from vertex $k$ in a turn.  Then, after the plays $(\ldots, X_y(1), a(1))$ any play deviating from the following sequences is a losing move: \\
$(b(1), e(1), f(1), d(1), g(1), X_z(1))$\\
 $(b(1), e(1), f(1), d(1), c(1), d(1), g(1), X_z(1))$.
\end{Lemma}
  
 We continue by analyzing all possible deviations from these sequences and show that they result in a loss.  In this claim, we will refer to the deviating player as the foe and the other player as the hero.  We will show that the foe loses in each case.  It may be helpful to refer to Figure \ref{fig:safeSequence} during these case descriptions.

\begin{itemize}
\item[1.]  \textbf{$c(1)$ instead of $e(1)$.}  Here we have two subcases: either moving from $g$ to $X_z$ is a winning (result is a $\fancy{P}$-position) or losing (a $\fancy{N}$-position) move.  If it's in $\fancy{N}$, then the hero can respond to $c(1)$ with $d(1)$.  If the foe then chooses $g(1)$, the hero can take the remaining stick in $d$ with $d(1)$.  $f(1)$ and $e(1)$ must follow with the hero winning.  If the foe instead chooses $f(1)$, the hero can win instantly by choosing $e(1)$.  For the foe's last chance, they could select $d(1)$, removing the other stick from $d$.  The hero should respond with $g(1)$.  The foe will lose by selecting $f(1)$, because the hero will win at $e(1)$, but the foe will also lose with $X_z(1)$, an $\fancy{N}$-position as assumed.
 
If $X_z(1)$ is instead leaves the board in $\fancy{P}$, the hero should respond to $c(1)$ with $d(2)$.  The foe could choose $f(1)$, but the hero can then win with $e(1)$.  Instead, the foe can choose $g(1)$ in which case the hero can choose $X_z(1)$ and win, as assumed.

\item[2.]  \textbf{$d(2)$ instead of (the first) $d(1)$.}  Here the hero has a simple move to win.  By taking $c(1)$ there are no further moves and the foe has lost.
  
\item[3.]  \textbf{$g(1)$ instead of (the first) $d(1)$.}  The hero can respond with $d(1)$.  This leaves two different adjacent vertices with 1 object apiece and no other adjacent non-empty vertices.  Either move by the foe results in one remaining move and a win for the hero.
 
\item[4.]  \textbf{$d(1)$ instead of $c(1)$.}  The hero can respond with $c(1)$ and win.

\item[5.]  \textbf{$d(1)$ instead of $X_z(1)$.}  This cannot happen in the second sequence, but if it happens in the first, the hero can respond with $c(1)$ and win.
\end{itemize}

Thus, any deviation from the two sequences specified in the claim puts the game in an $\fancy{N}$-position.  

\end{document}